\documentstyle[preprint,revtex]{aps}
\begin{document}
\draft
\preprint{to be submitted to Phys. Rev. C.}
\begin{title}
High--energy pion--nucleus elastic scattering
\end{title}

\author{C. M. Chen}
\begin{instit}
Los Alamos National Laboratory, Los Alamos, NM 87545 \\
and \\
Department of Physics and Center for Theoretical Physics,\\
Texas A\&M University, College Station, TX 77843
\end{instit}
\author{D. J. Ernst}
\begin{instit}
Department of Physics and Center for Theoretical Physics,\\
Texas A\&M University, College Station, TX 77843 \\
and \\
Department of Physics and Astronomy,
Vanderbilt University, Nashville, TN 37235
\end{instit}
\author{Mikkel B. Johnson}
\begin{instit}
Los Alamos National Laboratory, Los Alamos, NM 87545
\end{instit}
\receipt{21 December 1992}
\begin{abstract}
We investigate theoretical approaches to pion--nucleus elastic scattering at
high energies (300 $\le T_\pi \le$ 1 GeV). A ``model--exact'' calculation
of the lowest--order microscopic optical model, carried out in momentum space
and including the full Fermi averaging integration, a realistic off--shell
pion--nucleon scattering amplitude and fully covariant kinematics, is used to
calibrate a much simpler theory. The simpler theory
utilizes a local optical
potential with an eikonal propagator and includes the Coulomb interaction
and the first Wallace correction, both of which are found to be important.
Comparisons of differential cross sections out to beyond the second minimum
are made for light and heavy nuclei. Particularly for nuclei as heavy as
$^{40}$Ca, the eikonal theory is found to be an excellent approximation to
the full theory.
\end{abstract}
\narrowtext
\section{INTRODUCTION}
\label{secz:intro}
The pion--nucleus interaction on and below the $\Delta_{33}$ resonance
has been extensively studied, and microscopic models of elastic scattering
do a reasonable job \cite{john92} of describing the data.
The pion--nucleus interaction at energies above the $\Delta_{33}$ resonance
has received \cite{rokn88,will89,marl84,hipi} much less attention. In this
energy region (300 MeV $\le T_\pi \le $ 1 GeV), the pion has a much shorter
wavelength. For example, the wavelength at resonance is about 4 fm (about
the size of the nucleus)
while at 1 GeV the wavelength is 1 fm (about the size of a single nucleon).
The shorter wavelength implies that elastic and inelastic data at the
higher energies will provide sensitivity to the details of the spatial
dependence of the ground--state and transition densities. The data may also
prove sensitive to modifications \cite{sieg84,brow88} of the properties of
the nucleon in the nuclear medium, or reveal significant contributions from
exchange currents \cite{jiang92}.

Moreover, the pion--nucleon two--body interaction becomes much weaker as
one goes to energies above the $\Delta_{33}$ resonance.
The two--body total cross section becomes less than 30 mb, which is about 15
\% of that on the $\Delta_{33}$ resonance. The first implication of this
weaker amplitude is that the pion is able to penetrate deeper into the
nucleus. A simple estimate \cite{erns79} from total reaction cross
section studies shows that a projectile can penetrate into a target to
a radius that is equal to the impact parameter at which the profile
function equals one mean free path. In Fig.\ \ref{fig1} the profile
function for $^{12}$C, $^{40}$Ca and $^{208}$Pb are pictured, and the
arrows indicate approximately how far into a nucleus a pion of the
labeled energy can penetrate. The pion in the energy region from 500
MeV to 1 GeV is one of the most penetrating of the strongly interacting
particles.

The second implication of the weaker two-body cross section is
that multiple scattering theory for the optical potential becomes
increasingly convergent. A simple estimate of the convergence
is obtained by comparing a typical second--order term in the optical
potential to the first--order term.  For the case of short--range
correlations, one obtains \cite{erns77}
\begin{equation}
R\equiv {U^{(2)}\over U^{(1)}}=\sqrt{\sigma}{\ell_c\over k}\rho\,\,\,,
\end{equation}
where $\sigma$ is the total two--body cross section, $\ell_c$ is the
correlation length, and $\rho$ is the nuclear density at which the
pion interacts (See Fig. 1). The
factor of $1/k$, where $k$ is the incident pion momentum, reflects the
suppression of the pion propagator in $U^{(2)}$ with increasing energy.  On
resonance, we find $R\approx 0.1$, with $R < 1$ only
because $\rho$ is so small there.  At 500 MeV, we
find $R\approx 0.04$ and at 1 GeV, $R\approx 0.02$.
Thus differences between the data and a carefully calculated result
utilizing a first--order optical potential would be a strong indication
of the presence of unconventional phenomena such as mesonic current
contributions, modified nucleon properties in the medium, or other yet to be
thought of effects.

One option for calculating high--energy
pion scattering is a momentum--space optical--potential approach.  This
presents the
opportunity \cite{gieb88} to
calculate the scattering from a lowest--order optical potential
including the following features 1.) exact
Fermi--averaging integration, 2.) fully covariant kinematics \cite{erns80},
normalizations, and phase--space factors, 3.) invariant
amplitudes \cite{gieb88,erns90} and 4.) finite--range, physically motivated
two--body off--shell \cite{erns90} amplitudes. Within the multiple
scattering theory developed in Refs.\ \cite{john92,john83} this is an exact
calculation of the lowest--order optical potential.  A brief review of
this approach will be given in Sec.\ \ref{sec:formal}.

A practical difficulty arises when one wants to use the momentum--space
optical model to study high--energy pion--nucleus scattering. As the
projectile energy increases, more partial waves are needed in the
pion--nucleon
two--body amplitudes. Below 300 MeV, only S--waves and P--waves are needed.
Above 300 MeV, D--waves become important. F--waves become significant
above 500 MeV and G--waves and H--waves above 700 MeV. At the same time
the number of pion--nucleus partial waves is increasing at a rate
proportional to the pion momentum. At the high energies the momentum--space
approach becomes prohibitively computer--intensive, and one would like to
search for a simpler alternative.

The semi--classical theory immediately comes to mind as an alternative to
the momentum--space approach.  It is often used at high energies not only
because it is easier to compute, but also because the simpler character
of the theory facilitates obtaining physical insights into the reaction.
However, the semi--classical theory is
expected to be a good approximation only for local potentials and only when
the wavelength is sufficiently short.  Because the exact pion--nucleus
optical potential appears to be highly
nonlocal, we examine the role that nonlocalities play in
high--energy pion--nucleus scattering. This is done numerically in momentum
space and is presented in Sec. \ref{subsec:opt}.  We undertake this study
as a first step in obtaining a more quantitative measure of the validity
of the semi--classical theory than currently exist in the literature.

In Sec. \ref{subsec:eikon} we present our semi--classical model \cite{semi}.
In
order to have a quantitative as well as a simple model for confronting data, we
include the Coulomb interaction and the first Wallace correction
\cite{wallace}.
We establish the validity of the semi--classical model both by examining
the size of the Wallace correction and by comparing it to the model--exact
results obtained in momentum space.  Comparisons of the eikonal model to
model--exact calculations are made in Sec.\ \ref{sec:results}.

A previous similar investigation can be found in Ref.\ \cite{seki}.
There a factorized approximation \cite{pipit} was used in the
momentum--space calculations, so that the full nonlocality of the
optical potential was not considered. Furthermore, Glauber multiple scattering
theory\ \cite{glauber} was proposed, whereas we will examine an even
simpler model, that of a local optical potential with the scattering
solved via an eikonal approximation. A more detailed discussion that
compares and contrasts our results with previous work is given in
Sec.\ \ref{sec:comp}.
A summary, conclusions and future prospects are presented in Sec.\
\ref{sec:conc}.
\section{SCATTERING THEORY}
\label{sec:formal}
At the higher--energies where the pion--nucleon amplitude
becomes a smooth function of energy and is much weaker than on resonance,
the first--order optical potential may well be adequate for describing
all of the conventional nuclear physics phenomena that enter into the
reaction dynamics. Whether this is so remains for further investigation.
Independent of this question, there is the question of whether simpler
approaches to the calculation of the lowest--order optical potential
and the scattering from this potential are reasonably quantitative. It is
this question upon which we concentrate here. Below we first review the
momentum--space optical potential which serves as our `model--exact'
calculation. In momentum--space, we examine the importance of the
nonlocalities caused by both the resonance propagation and the finite
range of the two--body amplitude. Finding that these both can be accurately
approximated, we present a simple eikonal model.

Both the optical and the eikonal models use the same target
wave functions, which are obtained from Hartree--Fock calculations\
\cite{hf,bein75} with spurious center--of--mass motion removed as discussed in
\ref{sec:dcm}.  They also use the same on--shell pion--nucleon two--body
amplitudes,
which are constructed from Arndt's\ \cite{arndt} and H\"ohler's\
\cite{hohler} phase--shifts.
\subsection{Optical potential}
\label{subsec:opt}
A complete description of the momentum--space optical potential that we use
can be found in Ref.\ \cite{gieb88}, and a detailed discussion of
the covariant kinematics that we use can be found in Refs.\ \cite{erns80}.
The formal multiple--scattering theory in which this work is embedded is
given in Refs.\ \cite{john92,john83}. Here we provide only a brief
overview.

The first--order optical potential in the impulse approximation can be
written as,
\begin{eqnarray}
\langle \vec k'_\pi\vec k'_A\,\vert\,U_{\rm mo}(E)\,\vert\,
\vec k_\pi\vec k_A\rangle
=\sum_\alpha \int&&\,{d^3 k_{A-1}\over 2\bar E_{A-1}}\,{d^3 k'_N\over 2\bar
E'_N}\,{d^3 k_N\over 2\bar E_N}\,
\langle\Psi'_{\alpha ,\vec
k'_A}\,\vert\,\vec k'_N\vec k_{A-1}\rangle\nonumber \\
&&~~~~~~\times\,\langle \vec k'_\pi
\vec k'_N\,\vert\,t(E)\,\vert\,\vec k_\pi \vec k_N \rangle\langle \vec
k_N\vec k_{A-1}\,\vert\,\Psi_{\alpha ,\vec k_A}\rangle\,\,,
\label{eq:optical}
\end{eqnarray}
where $\vec k_\pi$, $\vec k_N$ and $\vec k_{A-1}$ are the momenta of the
pion, the struck nucleon and the $A-1$ residual nucleons in the
pion--nucleus center--of--momentum frame
respectively. $E$ is the incident energy of the pion and the nucleus
(in the pion--nucleus center--of--momentum frame also), and $\alpha$
is a set of quantum numbers that specifically label the nuclear bound state.

The target wave function $\langle \vec k_N\vec k_{A-1} \,\vert\,
\Psi_{\alpha ,\vec k_A}\rangle$ defined covariantly \cite{gieb88} contains
a momentum--conserving delta function $\delta (\vec k_A - \vec k_N - \vec
k_{A-1})$ and is a function of the relative momentum between the nucleon
and the $A-1$ residual nucleons. The pion--nucleon t-matrix similarly
contains a momentum--conserving delta function and is a function
of the relative momentum between the pion and the nucleon, $\vec \kappa$,
\begin{equation}
\langle \vec k'_N \vec k'_\pi \,\vert\,t(E) \,\vert\, \vec k_N \vec
k_\pi \rangle =
\delta (\vec k'_N + \vec k'_\pi - \vec k_N-\vec k_\pi)
\langle \vec \kappa' \,\vert\, t[\omega_\alpha (E,\vec k_\pi
,\vec k_N )] \,\vert\, \vec \kappa \rangle\,.
\end{equation}
One of the these three momentum--conserving delta functions leads to overall
momentum
conservation; the others allow one to perform two of the three
integrals in Eq.~\ref{eq:optical}. The remaining integral is the Fermi
averaging integration which
must be performed numerically. The details of how we perform the integration
can be found in \cite{gieb88}.

The energy at which one evaluates the two--body t--matrix, $\omega_\alpha
(E, \vec k_\pi, \vec k_N )$, must be carefully chosen \cite{erns85} if a
convergent perturbation theory is to result for calculations at and below the
$\Delta{33}$ resonance region. The energy $\omega_\alpha$ is defined
covariantly by
first defining the energy available to the pion--nucleon two--body subsystem
\begin{equation}
E_{\pi N}=E -\sqrt{(\vec k_\pi+\vec k_N)^2+m_{A-1}^2}\,\,,
\end{equation}
and then defining the invariant center-of-momentum energy for this system
\begin{equation}
\omega_\alpha^2=E_{\pi N}^2-(\vec k_\pi+\vec k_N)^2\,\,.
\label{eq:cmen}
\end{equation}
The mass of the $A-1$ system, $m_{A-1}$, differs from the mass of the
$A$--body target, $m_A$, by a nucleon mass and a binding energy,
$m_A=m_{A-1}+m_N+E_b$. This energy must then be shifted by a `mean-spectral
energy', $E_{ms}$, which is a calculated number \cite{erns85} that
approximately accounts for the interaction of the intermediate $\Delta_{33}$
(or the intermediate nucleon or other hadronic resonance) with the residual
nucleus. This first--order potential produces results for energies
near the $\Delta_{33}$ resonance that are in remarkable agreement with the
data. This is because \cite{john92} the sum of the second--order
effects (Pauli exclusion, true absorption and correlation corrections)
cancel amongst themselves. The use of invariant normalizations,
invariant phase
space and invariant amplitudes produces the phase-space factors that are
present in Eq. \ref{eq:optical}.

The optical potential so defined is then inserted into the Klein--Gordon
equation which is solved numerically to produce our model--exact calculation.

\subsection{Eikonal model}
\label{subsec:eikon}
The eikonal model we propose to examine results from first replacing
the optical potential of Eq.\ \ref{eq:optical} by a local potential and
then solving for the scattering amplitude arising from the use of the local
potential and the eikonal propagator. In order to include the Coulomb
interaction, we divide the scattering amplitude as
\begin{equation}
F(q)=F_{pt}(q)+F_{CN}(q)
\end{equation}
where $q$ is the momentum transferred to the pion, $\vec q=\vec k'_\pi-\vec
k_\pi$, $F_{pt}$ is the point Coulomb scattering amplitude,
and $F_{CN}$ is the scattering amplitude calculated from the sum of the
strong nuclear
potential plus a Coulomb correction, which is the difference between the
point Coulomb interaction and the Coulomb interaction of a finite charge
distribution. The eikonal approximation gives for $F_{CN}(q)$
\begin{equation}
F_{CN}(q)=ik_\pi \int_0^\infty
bdbJ_0(qb)e^{i\chi_{pt}(b)}\Gamma_{CN}(E,b),
\end{equation}
where $k_\pi$ is the momentum of the incident pion in the
pion--nucleus center--of--momentum frame, $b$ is the impact parameter,
$\chi_{pt}$ is the point Coulomb phase (in the eikonal approximation), and
$\Gamma_{CN}$ is the profile function defined below.

In order to incorporate the distortion and energy shift caused by the
Coulomb interaction, we follow the prescription given in Ref.\ \cite{coul}
and write the profile function as
\begin{eqnarray}
\Gamma_{CN}(E,b)&&=1-{\rm exp}\lbrace
i\chi_{CN}(E,b)\rbrace\nonumber \\
&&=1-{\rm exp}\lbrace i\chi_N\lbrack E,
b(1+EV_c(b)/k_\pi^2)\rbrack
+i\chi_c(b)-i\chi_{pt}(b)\rbrace,
\end{eqnarray}
where $V_c$ is the Coulomb potential of a uniformly
charged sphere,
$\chi_c$ is the phase shift caused by that potential, and the nuclear phase
shift $\chi_N$ is written as
\begin{equation}
\chi_N(E,b)=-{1\over 2k_\pi}\,\int_{-\infty}^\infty dz\,
U_{\rm eik}\lbrack E-V_c(r),r\rbrack,
\end{equation}
where
\begin{equation}
r=(b^2+z^2)^{1/2},
\end{equation}
and $U_{\rm eik}$ is the strong potential obtained as follows.
First we construct a local potential $U_o$ from
the pion--nucleon two--body scattering amplitude and the target density by
\begin{equation}
U_o(E,r)= -4\pi Z\lbrack f_p(0)\rho_p +{f'_p(0)\over
2k_\pi^2}\nabla^2\rho_p\rbrack
-4\pi N\lbrack f_n(0)\rho_n+{f'_n(0)\over
2k_\pi^2}\nabla^2\rho_n\rbrack,
\label{eq:eiku}
\end{equation}
where $Z$ and $N$ are the proton number and the neutron number
respectively. The scattering amplitude $f_p(0)$ [$f_n(0)$] is the
pion--proton [pion--neutron] scattering amplitude in the forward direction
in the pion--nucleus center--of--momentum frame. The finite range of the
pion--nucleon interaction is included through lowest nonvanishing order
and produces the additional terms in Eq.\ \ref{eq:eiku} which are
proportional to the derivatives of the scattering amplitude as shown in
Appendix \ref{sec:dir}.

In order to investigate the importance of the lowest--order corrections to
the eikonal propagator as derived by Wallace \cite{wallace}, next we
replace the potential $U_o (E,r)$ by the potential
\begin{equation}
U_{\rm eik}(E,r)=U_o+{U^2_o\over 4k_\pi^2}\Big( 1+{2b^2\over r}{d\over dr}
{\rm ln}U_o\Big).
\label{eq:wallace}
\end{equation}
The details of the derivation of this correction and higher order
corrections can be found in Ref.~\cite{wallace}.
\section{ RESULTS}
\label{sec:results}
In this section we investigate the importance of some of the ingredients of
our model--exact calculation of high--energy pion--nucleus scattering. We
will then examine how well our simple eikonal theory
reproduces the results of the model--exact theory.

\subsection{Contributions from nonlocalities in the optical potential}
\label{subsec:nonloc}
The most difficult part of evaluating the optical potential in the
model--exact theory is performing the Fermi--averaging integration.  This
is known to be very important in the region of the $\Delta_{33}$ resonance,
where proper accounting for the recoil and propagation of the pion--nucleus
resonance requires a careful treatment of the dependence of the
energy $\omega_\alpha$ in Eq.~\ref{eq:cmen} on the momentum of the nucleon.
Resonance propagation manifests itself as a nonlocality in the optical
potential, and this represents the first source of nonlocality whose
importance we want to assess for high--energy pions.

To investigate this we calculate elastic scattering of $\pi^-$ from
$^{40}$Ca  first with full Fermi averaging and then
utilizing a closure approximation.
The closure approximation that we use results from taking
\begin{equation}
\omega_\alpha^2= W_{\pi N}^2-k_\pi^2\,\,\,{\rm and}\,\,\,W_{\pi
N}=E-\sqrt{k_\pi^2+m_{A-1}^2}
\label{eq:closure}
\end{equation}
This is equivalent to setting the momentum of
the struck nucleon to zero in the pion--nucleus c.m. frame.

On the delta resonance, it has been found \cite{dhole,liurev,erns83}
that the closure approximation is totally
inadequate and even the `optimally-factorized' approximation \cite{erns83}
is not quantitative. At these higher energies the
conclusion is not {\it \'a priori} clear. The pion--nucleon amplitude
is, partial wave by partial wave, rather energy dependent as the individual
partial waves are resonant. The total amplitude, however, exhibits
only two very broad smooth peaks on a large background.

The resulting differential cross sections for pions at 500 MeV scattering
from $^{40}$Ca are shown in Fig.\ \ref{fig2}. We see that the correct
treatment of the recoil of the two--body pion--nucleon system only affects
the depth of the minimum but not to a very significant degree.

The behavior of the pion--nucleon scattering amplitude off--shell is taken
from the doorway--resonance model of Ref.\ \cite{erns90}.
Amplitudes which are resonant are separable in their dependence on the
relative momenta, hence maximally nonlocal. This is the second
source of nonlocality whose importance we want to investigate.  In the
doorway--resonance model, in each angular momentum, spin and isospin channel,
\begin{equation}
\langle \kappa'\,\vert\,t_{JLI} (\omega_\alpha )\,\vert\, \kappa
\rangle={v(\kappa')\over v(\kappa_o)}\,\langle\kappa_o
\,\vert\, t_{JLI} (\omega_\alpha )\,\vert\,\kappa_o\rangle\,
{v(\kappa)\over v(\kappa_o)},
\end{equation}
where $\kappa_o$ is the on--shell momentum, i.e. the momentum
corresponding to the center--of--momentum energy $\omega_\alpha $.
This nonlocality, which is also factored in the delta--hole \cite{dhole}
model, was found quite
important\cite{erns83} for energies on and below the $\Delta_{33}$
resonance.  For numerical convenience, we take the form factor to be a
Gaussian, $v(\kappa )=\exp (-\kappa^2/\beta^2)$, and we vary $\beta$ from
500 MeV to 4 GeV to investigate the importance of this nonlocality.  The
differential cross sections for $\pi^-$--$^{40}$Ca scattering calculated
with this variation in the form--factor range are shown in
Fig. \ref{fig3}. We see that the dependence on the off--shell range is
weak at this energy.

There exists some confusion in the literature concerning the
possible values for $\beta$. First, if one does not include
invariant phase space factors explicitly, they will appear effectively in
what one might wish to call the form factor. This difference is then one of
semantics. However, in deciding what might be a reasonable range over which
to vary the form factor, one should treat the kinematic factors explicitly.
In the $P$--wave channels, an artificial and incorrect increase
(in coordinate space) in the range of the form factor will also result if
the pion--nucleon pole term \cite{erns78} is not included in the model of
the pion--nucleon amplitude. For our purposes here, the difference of
interest is in examining  how the cross section changes in going from the
physically motivated doorway--model form factors to the higher momentum
cutoff. The 4 GeV cutoff produces an  approximately  zero--range interaction
in coordinate space. Figure \ref{fig3} demonstrates that the nonlocalities
contained in a finite range interaction do not alter significantly the
predicted elastic cross sections for these high energy pions.

\label{subsec:comp}
We have found that the nonlocalities in the optical potential at the high
energies do not have an important effect. This suggests that the simple
eikonal approach outlined above could prove to be an adequate model
for quantitative work at these energies. To investigate this, we
picture the differential cross sections predicted by each model
for $\pi^\pm$--$^{12}$C and $\pi^\pm$--$^{40}$Ca at 800 MeV/c in
Fig.\ \ref{fig4} and Fig.\ \ref{fig5}. The data are from Ref.\
\cite{marl84}. As stated earlier, we construct the target density from the
same wave
functions as were used in the momentum--space calculation and use the
same on--shell pion--nucleon scattering amplitude. For $^{40}$Ca, both the
location of the minima and the
magnitude of the cross section at the forward angles are in good agreement.
The agreement is, however, not as quantitative for
$^{12}$C where the minima are slightly shifted. This difference is
presumably caused in part by the fact that the kinematics
of target recoil enter explicitly in the momentum--space formulation
but not in the eikonal approach.
The eikonal model is thus more quantitative for
the heavier nuclei. We are examining the eikonal model to see if it
can be further improved for the light nuclei.

\subsection{Important features in the eikonal approximation}
\label{subsec:imeik}
In this section we investigate the role of the Coulomb interaction
and the first Wallace correction, which are both necessary if one wants
to use the eikonal model to quantitatively approximate the results of the
solution of the model--exact theory.

First, we find that the Coulomb interaction plays an important role in
determining the depth of the minima in the differential cross section. To
see exactly how important Coulomb--nuclear interference is, we
compare the calculation of the $\pi^-$--$^{40}$Ca differential cross
section at a series of energies as shown in Fig.\ \ref{fig6} with and
without the Coulomb interaction. The Wallace correction is not included in
either the solid or the dashed cureves. We find that the first minimum
in the differential cross sections becomes very deep when the Coulomb
interaction is included, falling well below $10^{-2}$ mb/sr in the region
of 500 to 600 MeV. Without the Coulomb interaction, the deepest minimum is
much shallower and occurs at 780 MeV. The inclusion of the Coulomb
interaction in the theory is necessary if a quantitative comparison with the
data for $\pi^-$--$^{40}$Ca is to be made. Coulomb--nuclear interference is
destructive for $\pi^-$ and constructive for $\pi^+$, which produces a less
dramatic effect for $\pi^+$--$^{40}$Ca scattering in this energy region.

Wallace has shown in Ref.~\cite{wallace} that the semi-classical
approximation can be improved if higher order corrections (known as
``Wallace'' corrections) are included. We examine in Fig.~\ref{fig7} the
importance of the first Wallace correction, Eq.~\ref{eq:wallace}, for
$\pi^-$--$^{40}$Ca scattering. We see in Fig.~\ref{fig7}, in the absence of
the Coulomb interaction, that the Wallace correction is dramatic on
resonance, $T_\pi= 180$ MeV, where it fills in the deep minimum. At the
higher energies, the correction interferes constructively until about 700
MeV filling in the minima somewhat. At 780 MeV, however, the Wallace
correction interferes destructively with $U_o$ and produce
a much deeper minimum.

In Fig.~\ref{fig8} we turn on both the Coulomb interaction and the Wallace
correction, we find a fascinating interplay between the two. Notice that at
400 MeV, for example, there two corrections are out of phase and tend to
cancel. At 680 MeV, however, they are in phase and interfere destructively
with $U_o$ producing the very deep minimum. By 780 MeV they
are both out of phase with each other but in phase with the lowest order
term and thus fill in the minima. Both the Coulomb and the Wallace
correction are necessary if there is to be a deep minima at 680 MeV, and not
at the neighboring energies. We note that the data of Ref.~\cite{marl84} was
taken at a pion lab momentum of 800 MeV/c which is very near $T_\pi= 680$
MeV. The depth of the first minimum of the differential cross section for
$\pi^-$ scattering over the energy range 400 to 800 MeV should prove a very
sensitive test of the existence, or lack thereof, of higher order
corrections in the reaction dynamics.

 \section{COMPARISON TO OTHER WORK}
\label{sec:comp}
Other work on high--energy pion--nucleus elastic scattering, which utilizes
the full Glauber theory, can be found in Refs.\ \cite{seki,oset}.
In particular, a comparison between the Glauber theory and the optical
model on $\pi^\pm$--$^{12}$C has been made in Ref.\ \cite{seki} and the two
approaches
were found to be in good agreement at the higher energies. We earlier
noted that the
optical potential model there did not include the full Fermi averaging,
and it was not clear to what extent the agreement found
was caused by the approximate treatment of the
non--localities in the optical potential. Our work clarifies that.
We also note that the Glauber theory they used is considerably more
complicated than our eikonal model.  To our knowledge, the Wallace
corrections have not been included in any of the numerical tests
of the eikonal model in this energy region.  Hence, the important
interplay between the Wallace and Coulomb corrections that we find has not
been
previously noted.  We also found that the center--of--mass
corrections in the target wave functions are important for $^{12}$C but
not for $^{40}$Ca, which is in agreement with Ref.\ \cite{seki}. Our work
strongly
supports their conclusion that pion scattering at the higher energies becomes
theoretically and calculationally simpler to treat at a quantitative
level.
\section{ SUMMARY, CONCLUSIONS, AND FUTURE PROSPECTS}
\label{sec:conc}
We have compared the scattering from the fully microscopic and
nonlocal lowest--order optical potential to the scattering from a simple
local potential in the eikonal approximation. We have investigated
the importance of several features of the full model.
Within the framework of our
``model--exact'' optical model, these include Fermi
averaging and the off--shell behavior of the pion--nucleon
scattering amplitude.
For the eikonal theory, we examine the importance of the first Wallace
correction and find that this correction is responsible for
a noticeable improvement of the eikonal theory in comparison to the
``model--exact'' theory. In all cases, the Coulomb--nuclear interference
is important and must be included if one is to compare with data.

This semi--classical eikonal theory put forth here is very simple, even
simpler than the Glauber theory utilized by others.
It appears to be quantitatively
valid at high energy, at least for the first several minima in the
differential elastic cross section. This makes possible a much simpler
reaction theory than has been needed at energies at and below
the $\Delta_{33}$ resonance.

At high energies and/or for heavy nuclei, use of the full
momentum--space optical model becomes very time--consuming and sometimes
computationally impossible.
In contrast, the eikonal theory is relatively simple and fast on the
computer. It therefore becomes a matter of considerable practical
importance to realize that the physics is faithfully reproduced by the
simple version of the theory out to the position of the second minimum
in the differential cross section.

An examination of the pion--nucleon cross section shows the
complex interplay
of resonances. Ultimately, it is interesting to explore how resonances behave
in the nucleus and how the specific features of the optical potential
are suited for nuclear structure studies. In this pursuit, it will be very
helpful to
have a simple version of the theory as developed here. We will pursue these
questions in subsequent work.

The momentum space optical model has been extended to treat kaon-nucleus
scattering \cite{chen92}. Results show qualitative  agreement with previous
work \cite{sieg84}. Modifications to the  kaon-nucleon amplitudes
in the
nuclear medium are needed to eliminate discrepancies between the theory and
the data. On the quantitative level, results from the momentum space
approach give larger discrepancies between theory and data than do those
from the coordinate space
calculations used in Ref.\ \cite{sieg84}. Our eikonal model, which is also a
coordinate space approach and been shown to be a good approximation
to the ``model-exact'' momentum space theory for high-energy pion
scattering, can also be modified to treat the kaon scattering. Results from
the eikonal model may reveal the sources of the discrepancies between the
momentum space and the previous coordinate space calculations.
Work on the kaon-nucleus scattering utilizing the eikonal model is in
progress.
\newpage
\appendix{CENTER--OF--MASS CORRECTIONS}
\label{sec:dcm}
The scattering theory developed in Ref.~\cite{gieb88} is covariant. However,
to implement it covariantly would require a covariant model of the nucleus.
This would clearly be overkill as the nucleus for the cases of interest is
not moving relativistically. A Galilean invariant model would thus suffice.
Contemporary models of the nucleus, however, are not Galilean invariant.
We use Hartree--Fock models of the nucleus which contain spurious
center-of-mass motion. The approach we adopt is discussed in detail in
Ref.~\cite{erns80}. The Hartree--Fock results are fitted to elastic
electron scattering. The spurious center-of-mass motion is there treated
by requiring that momentum conservation holds for average values. We
make the same approximation for pion--nucleus scattering.

Specifically we calculate the target wave functions from the Hartree--Fock
model and then calculate the Fourier transform of the density. We require
the intrinsic density, i.e. the density without the center-of-mass motion.
If the wave function were a pure harmonic oscillator, then the correction
could be made exactly. Here, we make the correction as if the Hartree--Fock
result were exactly a harmonic oscillator. The result \cite{elli55} for
obtaining the intrinsic proton density $\rho^p_{cm}$ from the Hartree--Fock
density $\rho^p_{HF}$ is
\begin{equation}
\rho^p_{cm} (r)=\int\,d^3r_1\,(B\sqrt{\pi})^{-3}\,e^{(r_1^2/B^2)}\,
\rho^p_{HF}(\vert \vec r-\vec r_1\vert)\,\,\,,
\end{equation}
where $B=b/\sqrt{A}$ and $b$ is the usual harmonic oscillator parameter.
The Hartree-Fock calculation uses a slightly different $b$ for each orbital
so we use the average value in making the center--of--mass correction.

We use $\rho^p_{cm}$ to calculate the r.m.s.~radius and compare this to the
r.m.s.~radius measured in electron scattering (with the finite size of the
proton removed). For $^{12}$C and the Hartree--Fock result from
Ref.~\cite{bein75} we find that the r.m.s. radius is 2.365 fm from the
Hartree--Fock calculation while it is measured to be 2.297 fm in electron
scattering.
We therefore scale the $b$'s in the Hartree--Fock calculation by .9715 (a
3\% correction) so that the correct measured r.m.s.~radius is
reproduced. After the scaling, we calculate the predicted electron
scattering and find that these wave functions reproduce electron
scattering data out to a momentum transfer which is larger than is
encountered in pion scattering. For $^{40}$Ca, these corrections were
already made by Negele \cite{hf} and his wave functions are consistent
with the electron scattering data.
\appendix{FORWARD SCATTERING AMPLITUDE}
\label{sec:dir}
Using a Taylor expansion, we can write the forward scattering amplitude for
small angles as
\begin{eqnarray}
\label{eq:amplitude}
f(\theta)&\approx& f(0)+{df(\theta)\over d\cos
\theta}\Big\vert_{\cos\theta=1}(\cos\theta-1)\nonumber\\
&\approx& f(0)- f'(0){\theta^2\over 2},
\end{eqnarray}
where
\begin{eqnarray}
f'(0)&=&{df(\theta)\over
d\cos\theta}\Big\vert_{\cos\theta=1}\nonumber\\
&=&\sum_l f_l(0){dP_l(\cos\theta)\over
d\cos\theta}\Big\vert_{\cos\theta=1}\nonumber\\
&=&\sum_l f_l(0){l(l+1 )\over 2}.
\end{eqnarray}
We further use the approximation
\begin{equation}
\theta\approx {q\over k_\pi},
\end{equation}
and the fact
that the momentum transfer $q$ in momentum space corresponds to $i\nabla$ in
coordinate space, to rewrite  Eq.\ \ref{eq:amplitude} as
\begin{equation}
f(\theta)\approx f(0)+ {f'(0)\over 2k_\pi^2}\nabla^2.
\end{equation}
To construct a local optical potential, we multiply $f(\theta)$ by the
target density $\rho$ and obtain Eq.\ \ref{eq:eiku}.

The forward scattering amplitude $f(0)$ and its derivative $f'(0)$ are
converted from phase shifts in the pion--nucleon c.m. frame, then transferred
to the pion-nucleus c.m. frame through the relation:
\begin{equation}
f(0)={k_\pi\over \kappa}f_{\pi N}(0),
\end{equation}
where $\kappa$ and $f_{\pi N}(0)$ denote the pion momentum and forward
scattering amplitude in the pion--nucleon c.m. frame.

\figure{Profile function $S(b)$ for $^{12}$C, $^{40}$Ca and $^{208}$Pb.
The arrows indicate the depth to which the pion (labeled by its energy)
can penetrate.\label{fig1}}
\figure{The differential cross section for elastic scattering of $\pi^-$
from $^{40}$Ca at 500 MeV verses center-of-mass angle. The solid curve
is the result of a calculation in which the full Fermi averaging is
performed exactly, while the dashed curve uses the closure
approximation defined in Eq. \ref{eq:closure}.
\label{fig2}}
\figure{The differential cross section calculated from different form
factor ranges $\beta$ defined in Sec.\ref{sec:results}  for
$\pi^-$--$^{40}$Ca
elastic scattering at 500 MeV. The solid curve is from the ``model--exact''
theory. The doted curve corresponds to $\beta=$ 4 GeV. The dashed curve is
the result from setting $\beta=500$ MeV.
\label{fig3}}
\figure{The elastic differential cross sections for $\pi^\pm$--$^{12}$C
scattering at a pion lab momentum 800 MeV/c. The solid curves are from the
momentum space calculations,
while the dashed curves are from the eikonal model. Data are obtained from
Ref. \cite{marl84}.
 \label{fig4}}
\figure{The same as Fig.\ref{fig4} except the target is $^{40}$Ca.
\label{fig5}}
\figure{$\pi^-$--$^{40}$Ca elastic differential cross section calculated
from the eikonal model at labeled pion lab energies. The Coulomb interaction
is included in all the solid curves, but not the dashed curves. The Wallace
correction is not  included here.
  \label{fig6}}
\figure{$\pi^-$--$^{40}$Ca elastic differential cross sections
calculated from the eikonal model at labeled pion lab energies. The dashed
curves are the same as in Fig.~\ref{fig6}. The solid curves include the
Wallace correction but not the Coulomb interaction.
\label{fig7}}
\figure{The dashes curves are the same as in Figs.~\ref{fig6} and
\ref{fig7}. The solid curves now contain both the Wallace correction and the
Coulomb interaction.
\label{fig8}}
\end{document}